\documentstyle[amssymb,sprocl,epsfig]{article}

\def\Journal#1#2#3#4{{#1} {\bf #2}, #3 (#4)}

\def\be{\begin{equation}}
\def\ee{\end{equation}}
\def\bea{\begin{eqnarray}}
\def\eea{\end{eqnarray}}

\begin{document}

\title{X-RAY SPECTROSCOPY AND TIME VARIABILITY OF NARROW LINE SEYFERT 1}

\author{F. CANCELLIERE}

\address{Dipartimento di Astronomia, Universit\`a di Bologna,
via Ranzani 1, 40127 Bologna\\E-mail: frank@kennet.bo.astro.it}

\author{A. COMASTRI}

\address{INAF-Osservatorio Astronomico di Bologna, via Ranzani 1,
40127 Bologna\\E-mail: comastri@anastasia.bo.astro.it}


\maketitle\abstracts{ We present the main results of broad band (0.1-10 keV) BeppoSAX observations
of a selected sample of Narrow Line Seyfert 1 galaxies (NLS1s). We have performed a spectroscopic analysis of LECS and MECS data to test their average spectral properties, and timing analysis to study their amplitude variability.
The steep average spectral indices and the high amplitude time variability are
in agreement with a scenario where NLS1s, with respect to the Broad Line Seyfert 1 (BLS1s) host a smaller 
black hole mass surrounded by hotter accretion disc.}

\section{Introduction}
Narrow Line Seyferts 1 are a peculiar class of extragalactic objects, presenting several
interesting and complex characteristics. They are defined on the basis of the  nuclear optical spectrum with the following criteria: i) FWHM(H$\beta$) $\lesssim$2000 km s$^{-1}$; ii) [OIII]/H$\beta<$3; iii) narrow permitted lines only slightly broader than the forbidden ones (Osterbrock $\&$ Pogge 1985). In the X-ray band they are characterized by: i) steep spectral indices both in the soft (Boller et al. 1996) and hard (Brandt et al. 1997) band with $\Gamma$$\simeq$2$\div$5, compared to $\Gamma$$\simeq$ 1.8$\div$2 for BLS1; ii) rapid, high-amplitude X-ray variability (Boller et al. 1997); iii) strong soft X-ray excess (Pounds et al. 1995); iv) spectral features due to ionized iron (Comastri et al. 1998).
An observational program of a small
sample of these objects has been carried out using the LECS and MECS instruments onboard BeppoSAX satellite, with the
aim to investigate their broad band X-ray spectral and time variability properties and to compare theme with those of Seyfert galaxies with broad optical lines (BLS1). 
We used a sample of seven  well known NLS1s (PG 1115+407, RE J1034+396, ARK 564, Ton S180, IRAS 13224-3809, IRAS 13349+2438, PKS 0558-504) 
and, as a comparison sample, 10 bright BLS1s also observed with BeppoSAX (Fairall 9, IC 4329A, Mrk509, Mrk841, MCG 6-30-15, MCG 8-11-11, NGC 3783, NGC 4593, NGC 5548, NGC 7469, Matt 2000).

\section{Spectral properties}
The BeppoSAX X-ray spectra of both NLS1s and BLS1s have been fitted with a single power law plus Galactic absorption. Although this model is not always the best-fit solution (see Comastri 2000 for a summary of the spectral analysis results), it provides a first estimate of the average  shape of the X-ray spectrum. 
The mean values of the photon indices ($\Gamma$) plus dispersion in the soft (0.1-2 keV), hard (2-10 keV) and total (0.1-10 keV) band are:
\vspace{0.4cm}

$<\Gamma _{soft}>_{N}$ = 2.92 $\pm$ 0.63 \hspace{1.3cm} 
$<\Gamma _{soft}>_{B}$ = 1.90 $\pm$ 0.38

$<\Gamma _{hard}>_{N}$ = 2.11 $\pm$ 0.29 \hspace{1.3cm} 
$<\Gamma _{hard}>_{B}$ = 1.77 $\pm$ 0.15

$<\Gamma _{0.1-10}>_{N}$ = 2.14 $\pm$ 0.42 \hspace{1cm} 
$<\Gamma _{0.1-10}>_{B}$ = 1.93 $\pm$ 0.29
\vspace{0.4cm}


\subsection{Spectral analysis of IRAS 13349+2438}
The BeppoSAx spectrum of {\bf IRAS~13349+2438} requires in addition to a power law ($\Gamma = 1.98_{-0.15}^{+0.12}$) plus Galactic absorption, a warm absorber component ($N_{H}$=($1.44^{+1.01}_{-0.95}$)$\times10^{22}$ cm$^{-2}$, $\xi=91.7^{+101}_{-56}$ erg cm s$^{-1}$) and an absorption edge at E$\simeq$8 keV consistent with ionized iron. There is only a marginal evidence of a iron emission line around 6.7 keV (EW $<$ 225 eV).  
Ionized emission features are generally expected in NLS1s spectra 
(e.g. Comastri et al. 2001, Turner et al. 2001), often associated with absorption edges, suggesting a ionized emitting/absorbing medium, probably 
the surface layers of a highly ionized accretion disc (Matt et al. 1993). 

\section{Variability in NLS1s}

In order to quantify the amplitude variability, we followed the method developed
in Nandra et al. (1997). More specifically 
we calculated the normalized ``excess variance'', \( \sigma ^{2}_{rms} \),
of each light curve 

\[
\sigma ^{2}_{rms}=\frac{1}{N\mu ^{2}}\sum ^{N}_{i=1}[(X_{i}-\mu )^{2}-\sigma ^{2}_{i}]\]
 
The error on \( \sigma ^{2}_{rms} \) is given by s\( _{D} \)/(\( \mu ^{2} \)\( \sqrt{N} \)),
where 
\[
s^{2}_{D}=\frac{1}{N-1}\sum ^{N}_{i=1}\{[(X_{i}-\mu )^{2}-\sigma _{i})^{2}]-\sigma ^{2}_{rms}\mu ^{2}\}^{2}\]

where \emph{X}\( _{i} \) is the count rate for the N points in each
light curve, with errors \( \sigma _{i} \) and \( \mu  \)
is the unweighted arithmetic mean of the \emph{X}\( _{i} \). 

The average values of the excess variance (plus dispersion) in the soft (0.1-2 keV) and hard (2-10 keV) band are reported below (in units of $10^{-2}$).
\vspace{0.4cm}

$<\sigma_{soft}^{2}>_{N}$ = (4.571 $\pm$ 0.322) \hspace{0.32cm} 
$<\sigma_{soft}^{2}>_{B}$ = (1.320 $\pm$ 0.081)

$<\sigma_{hard}^{2}>_{N}$ = (6.292 $\pm$ 0.592) \hspace{0.3cm} 
$<\sigma_{hard}^{2}>_{B}$ = (0.504 $\pm$ 0.016) 
\vspace{0.4cm}

The excess variance in both energy bands is clearly higher for the NLS1s. 

\begin{figure}[h]
\begin{tabular}{cc}
\psfig{figure=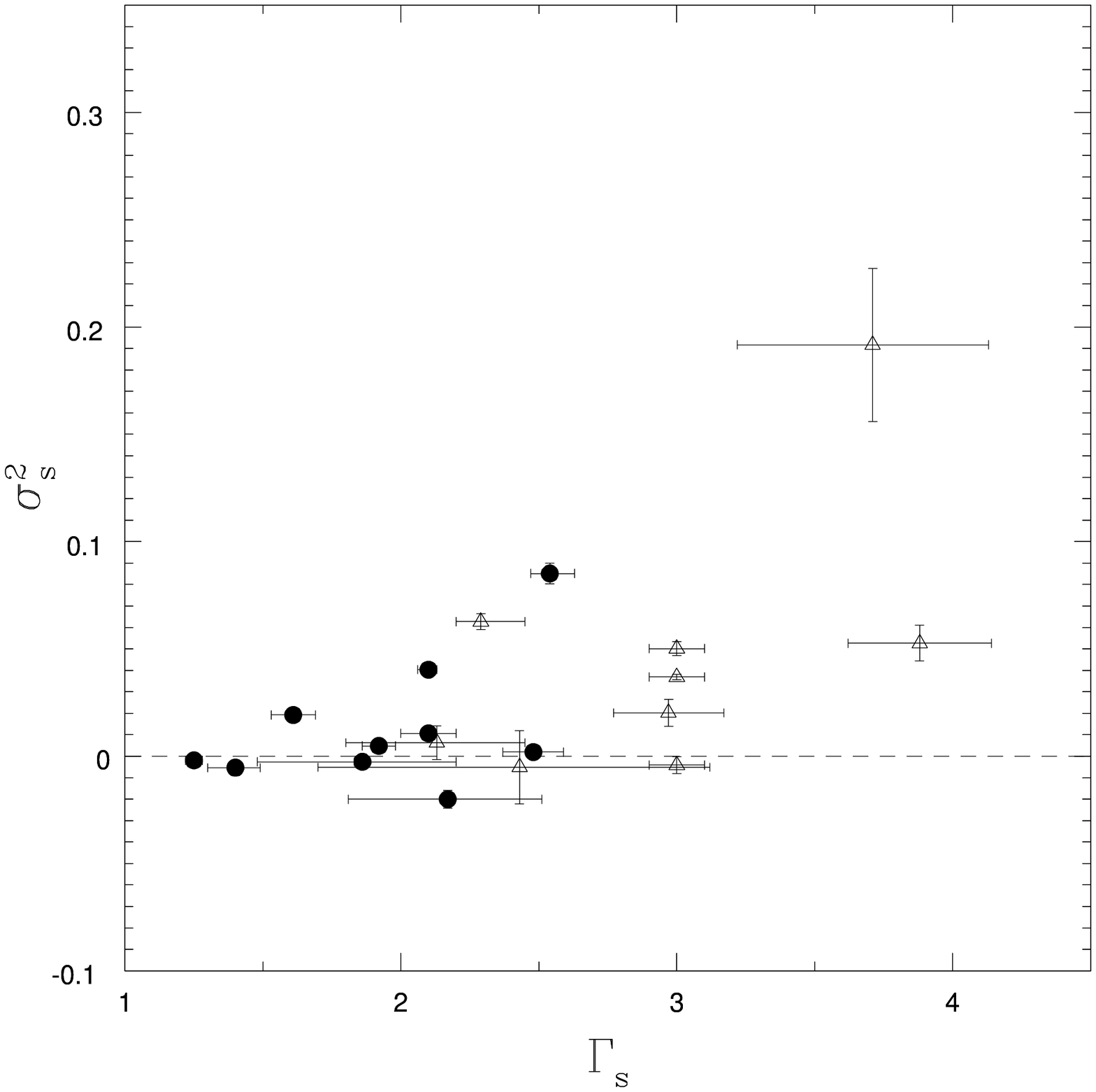, width=5cm, height=5cm}
&\psfig{figure=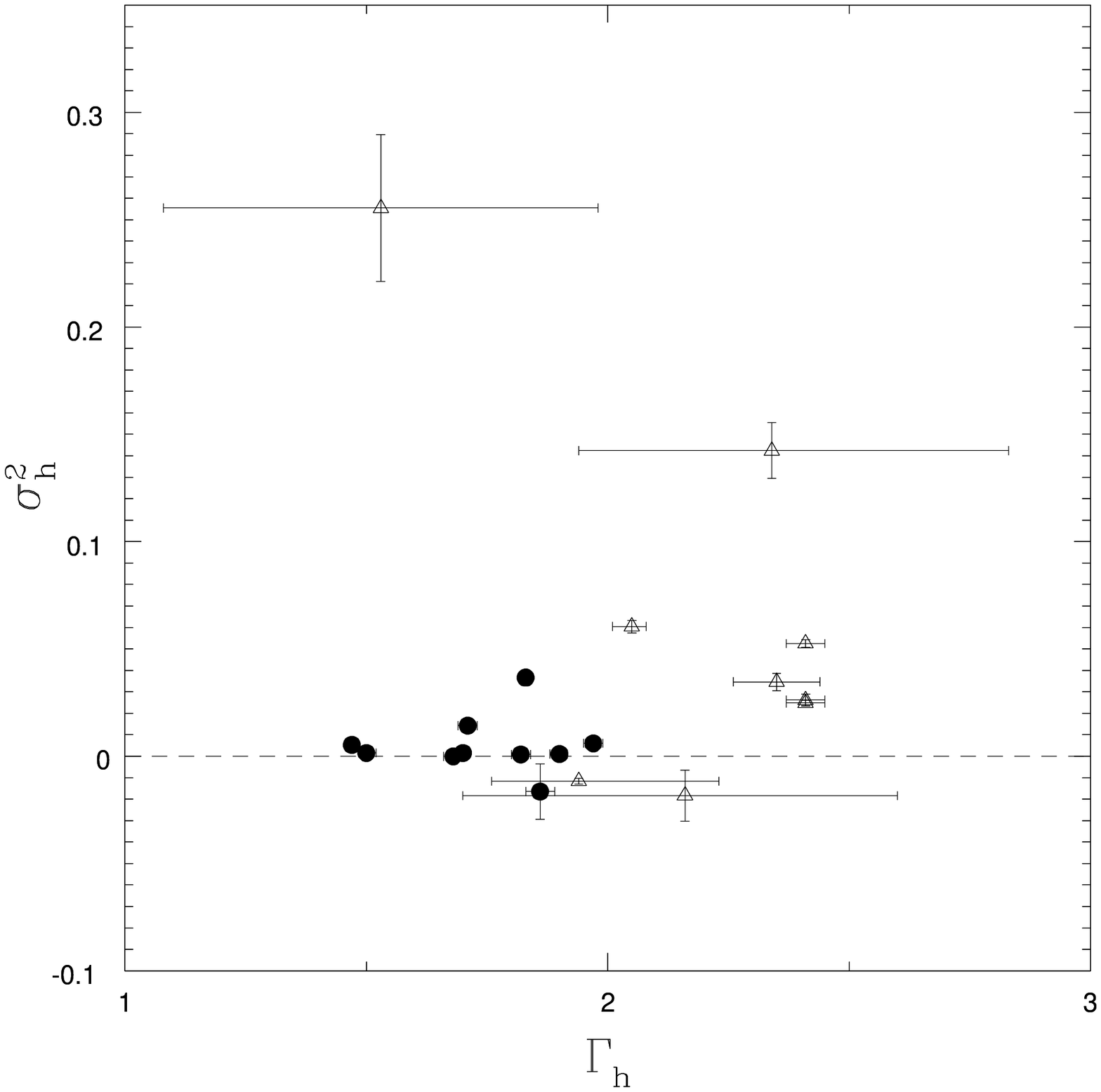, width=5cm, height=5cm} \\
\psfig{figure=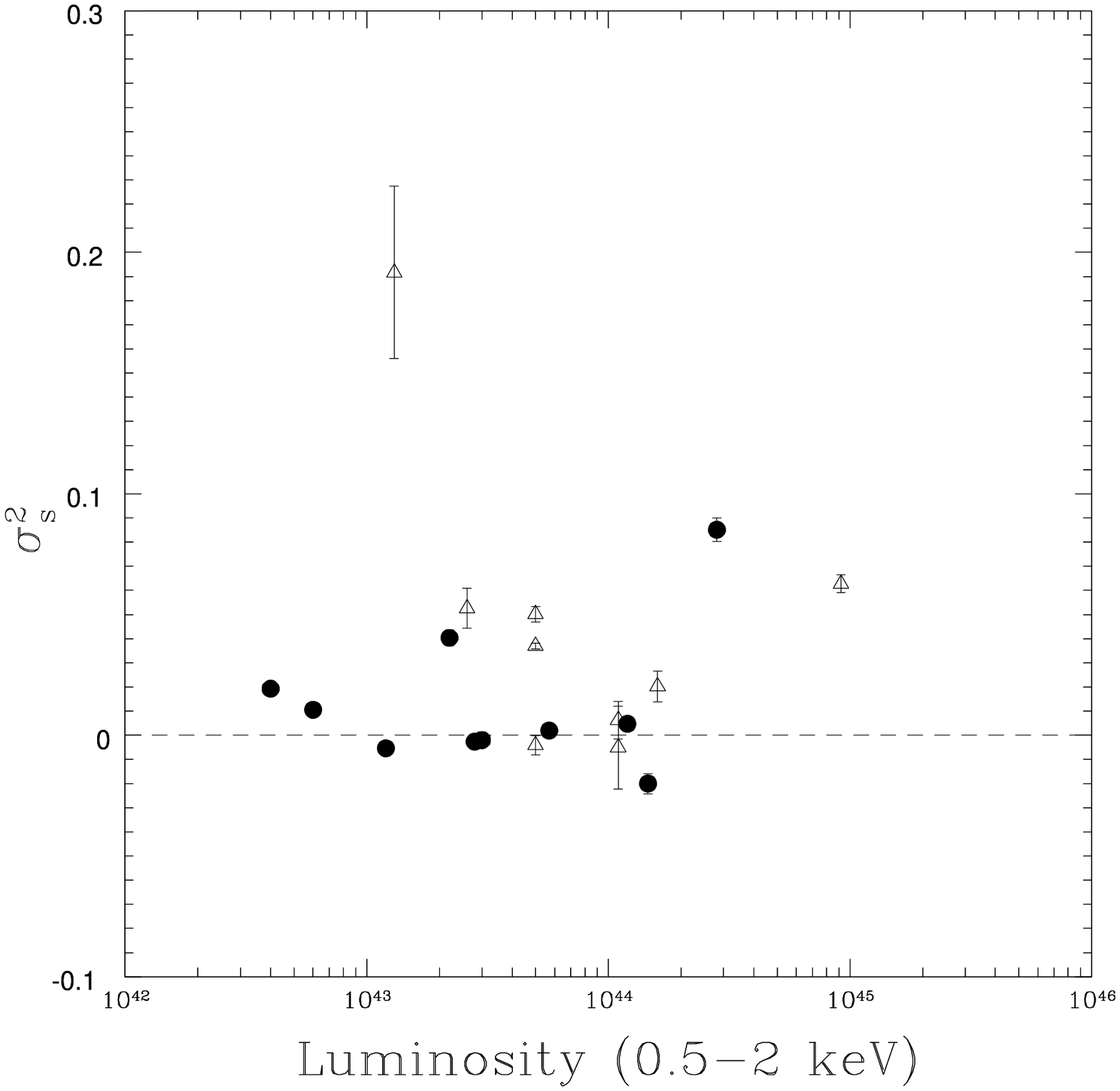, width=5cm, height=5cm}
&\psfig{figure=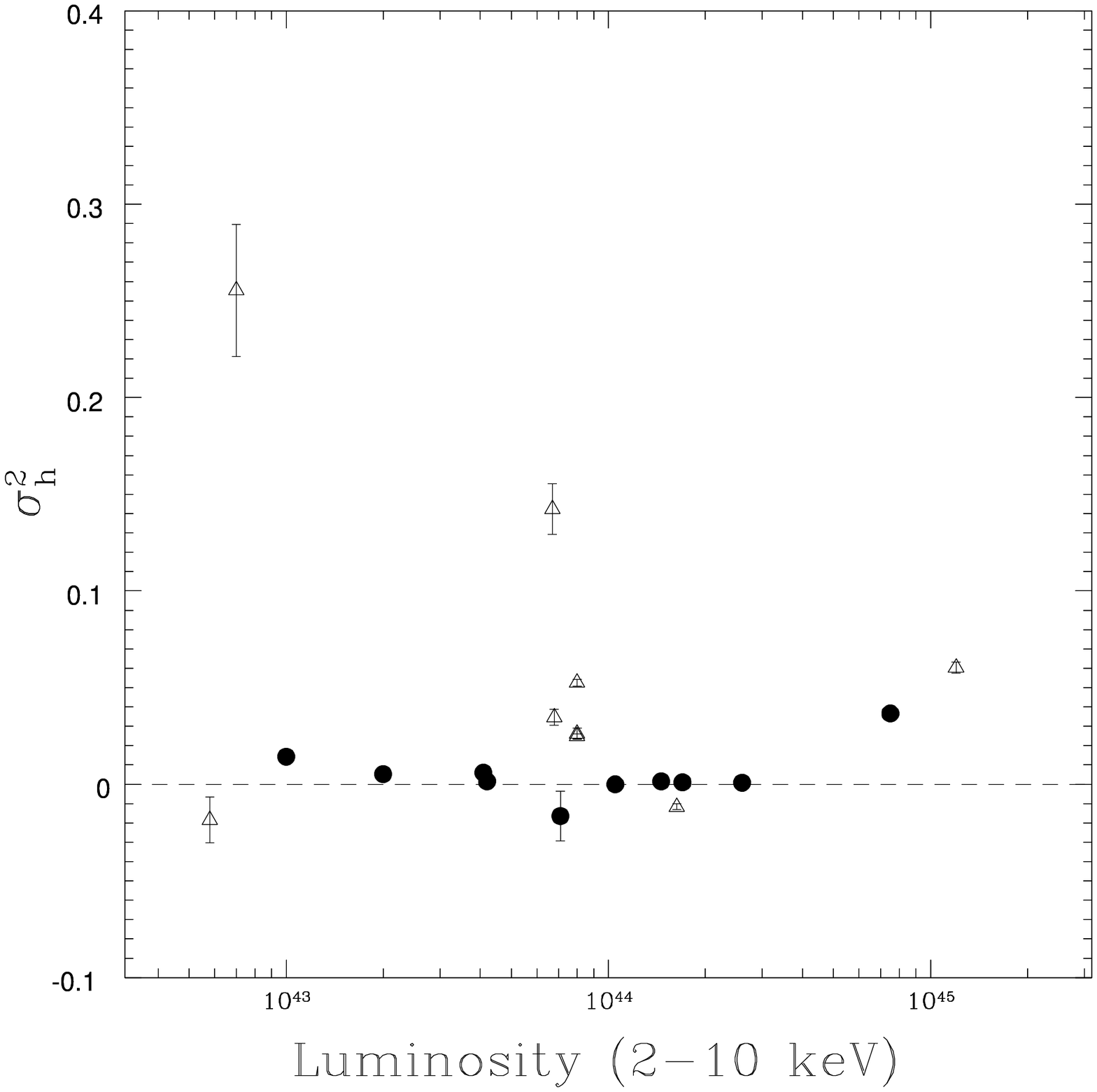, width=5cm, height=5cm} 
\end{tabular}
\caption{Upper panel-left: correlation between $\Gamma_{soft}$ and $\sigma^{2}_{soft}$; right: correlation between $\Gamma_{hard}$ and $\sigma^{2}_{hard}$.
Lower panel-left: correlation between soft luminosity and $\sigma^{2}_{soft}$; right: correlation between hard luminosity and $\sigma^{2}_{hard}$ (BLS1s: filled circles, NLS1s: triangles). }
\end{figure}

We have also searched for possible correlations between \( \sigma ^{2}_{rms} \), the  X-ray luminosity and the spectral index (fig. 1). 
The spectral index in both soft and hard bands seems to be correlated
to \( \sigma ^{2}_{rms} \) , while only a weak trend between the excess variance and the luminosity is present. These results, which need to be supported by further studies with a larger sample, appears to be inconsistent with the ASCA results (Leighly 1999), where the excess variance $\sigma^{2}_{rms}$ is anticorrelated with the  X-ray luminosity. 

\section{Conclusions}
From the spectral analysis of a sample of seven NLS1s observed by BeppoSAX, we confirm that they have steeper spectral indices and faster X-ray variability with respect to BLS1s. The present results fit with the idea that NLS1s are powered by a relatively small central black hole accreting near the Eddington limit.

\section*{Acknowledgments}
F.C. thanks Elisa Costantini, Piero Ranalli and Cristian Vignali for many helpful and useful discussions.
This research has been partially supported by ASI contracts I/R/113/01 and I/R/073/01.

\end{document}